\documentstyle[12pt]{article}
\makeatletter

\long\def\@caption#1[#2]#3{
  \begingroup
    \@parboxrestore
    \small
    \baselineskip=12pt
    \advance\leftskip by 1cm
    \advance\rightskip by 1cm
    \@makecaption{\csname fnum@#1\endcsname}{\ignorespaces
     {#3}}\par
  \endgroup}
\makeatother

\begin{document}
\baselineskip=24pt
\centerline {\large \bf Gamow-Teller Strength in the
Region of $^{100}$Sn}

\vspace{ 24pt}
\centerline{\bf B. Alex Brown}

\vspace{ 12pt}
\baselineskip=18pt
\centerline{\sl National Superconducting Cyclotron Laboratory}
\centerline{and}
\centerline{\sl Department of Physics and Astronomy,}
\centerline{\sl Michigan State University,}
\centerline{\sl East Lansing, Michigan 48824-1321, USA}

\vspace{ 12pt}
\centerline{ and }

\vspace{ 12pt}
\centerline{\bf K. Rykaczewski}

\vspace{ 12pt}
\baselineskip=18pt
\centerline{\sl Institute of Experimental Physics}
\centerline{\sl Warsaw University,}
\centerline{\sl Pl-00681 Warsaw, Hoza 69, Poland}

\vspace{ 36pt}
\baselineskip=24pt
\begin{myab}
\hspace{0.5cm}{\bf Abstract:}
New calculations are presented for Gamow-Teller beta decay of
nuclei near $^{100}$Sn. Essentially all of the
$^{100}$Sn Gamow-Teller decay strength is predicted to
go to a single state at an excitation energy of 1.8 MeV
in $^{100}$In. The first calculations are presented for
the decays of neighboring odd-even and odd-odd nuclei which
show, in contrast to $^{100}$Sn, surprisingly complex and broad
Gamow-Teller strength distributions. The results are compared
to existing experimental data and the resulting hindrance
factors are discussed.
\end{myab}

\vspace{ 36pt}
\centerline {{\bf PACS:} 21.10.Pc, 21.60.Cs, 23.40.-s,27.30.$+$t}

\clearpage
One of the primary new directions in nuclear spectroscopy is in the
experimental study and theoretical understanding of nuclei near the
limits of particle stability.
The heaviest nucleus with an equal number
of protons and neutrons which is predicted to be stable is $^{100}$Sn,
and
experiments are being planned and
carried out at several laboratories to produce and
study the decay of this nucleus\,$^{1}$
and others\,$^{2,3}$ in this mass region. One of
that most interesting aspects of these proton-rich nuclei is the
most of the giant Gamow-Teller resonance lies within the beta-decay
Q-value window. We report here on new calculations which show some of
the unusual features which one may expect to see in these decays, the
special problems associated with their experimental detection, and the
important nuclear
structure information which will be obtained.

Our model space, which is similar to that of a number
of other
calculations,\,$^{4,5,6,7}$
is designed for nuclei with Z$\leq$50 and N$\geq$50
and starts from a closed-shell configuration for $^{100}$Sn.
We will later discuss the effects of going beyond the
closed shell configuration. In the model space we designate
by SNA, proton holes are allowed to occupy the
0f$_{5/2}$, 1p$_{3/2}$, 1p$_{1/2}$ and 0g$_{9/2}$ orbitals, and the
neutron
particles occupy the 0g$_{7/2}$, 1d$_{5/2}$, 1d$_{3/2}$, 2s$_{1/2}$
and 0h$_{11/2}$ orbitals. The single-particle
energies (SPE) and two-body matrix elements (TBME) for the protons
in model space SNA are those of Ji and Wildenthal\,$^{8}$
which were obtained from a least-squares fit to energy
levels of the N=50 isotones. For the neutron residual
interaction, we started with a set of
TBME obtained from a similar least-squares fit to the N=82
isotones with a $^{132}$Sn\,$^{9}$ core
in which the protons fill the same set of
orbitals as do the neutrons outside of the $^{100}$Sn core.
We then subtracted a calculated Coulomb interaction
and scaled the resulting TBME by a factor of (132/100)$^{0.3}$.
The scaling approximately takes into account the change in
size of the valence wave functions between $^{132}$Sn and
$^{100}$Sn. The proton-neutron interaction was calculated from the
bare G matrix of Hosaka\,$^{10}$ which is
based on the Paris potential. Finally,
the neutron single-particle energies were determined from a
consideration of the ``single-particle" states observed for the
odd-even N=51 nuclei and will be discussed below.

We are interested in calculating the level
structure and decay properties for as many nuclei as possible
away from $^{100}$Sn. We are also constrained by
computational limitations to the consideration of J-T
Hamiltonian
matrix dimensions below about 10000. In model space SNA this
constraint limits the $\beta^{ + }$ decay calculations to those
initial nuclei with N$_{p}+$N$_{n}\leq$4, where N$_{p}$ are the number
of valence proton holes and N$_{n}$ are the number of valence
neutron particles.
To go to larger N$_{p}$ values, we investigated
model space SNB in which only the 1p$_{1/2}$ and 0g$_{9/2}$
proton orbitals are active. The interaction is, of course,
model-space dependent and we replace the Ji-Wildenthal
SPE and TBME with the seniority conserving interaction
Gloeckner and Serduke.\,$^{11}$
With these changes (and
keeping the neutron and proton-neutron parameters the same),
we recalculated the Gamow-Teller decay spectrum of $^{98}$Cd and
found it to be essentially the same as that obtained in the larger
SNA model space.
(This result disagrees with similar comparisons
made in Ref 4 and Ref 5. This is
related to the fact that the previous work did not take into account the
renormalization of the
proton-proton interaction on going from SNA to SNB.)

Finally, we come back to a discussion of
the neutron single-particle energies
and the related proton-neutron interaction which are particularly
important for the Gamow-Teller decay properties. The ground states of
all known odd-even isotones with N=51 from $^{89}$Sr to
$^{97}$Pd have J$^{ \pi }$ = 5/2$^{ + }$. One-neutron transfer reactions
on $^{88}$Sr and $^{90}$Zr establish these as 1d$_{5/2}$ single-particle
states and also provide information on the location of the
excited 0g$_{7/2}$, 1d$_{3/2}$ and 2s$_{1/2}$ states.\,$^{12}$ In
addition,
it is known that the excitation energy
of the 7/2$^{ + }$ states comes down linearly
from about 2.0 MeV in $^{91}$Zr to about 0.6 MeV in
$^{97}$Pd.\,$^{7}$ A reduction of the gap between the 0g$_{7/2}$ and
1d$_{5/2}$ single-particle states is obtained in the
SNB model-space due to the relatively large
proton-neutron TBME connecting the 0g$_{9/2}$ and 0g$_{7/2}$ orbitals.
However, the reduction compared to experiment
is too strong
by about 30\%. Better agreement can be
obtained by renormalizing the proton-neutron G matrix
elements by a factor of 0.7. This renormalization
improves agreement with experiment for the absolute
change in the neutron SPE between $^{89}$Sr and $^{97}$Pd,
and also improves the agreement with the location of the
strong GT states in the $\beta^{ + }$ decay of $^{98}$Cd. Thus,
we have adopted this renormalization for all calculations within
the SNB model spaces. The absolute single-particle
energies in units of MeV
relative to a $^{100}$Sn closed shell in model space
SNB are for protons $-$3.38 (1p$_{1/2}$) and $-$2.99 (0g$_{9/2}$), and
for neutrons
$-$10.15 (0g$_{7/2}$), $-$10.10 (1d$_{5/2}$), $-$8.09 (1d$_{3/2}$),
$-$8.40 (2s$_{1/2}$) and $-$7.85 (0h$_{11/2}$). It is interesting to
note the crossover of the 0g$_{7/2}$ and 1d$_{5/2}$ states
in $^{101}$Sn relative to the other N=51 nuclei, and it would
be very important to have an experimental
confirmation of the ground state spin and level structure of $^{101}$Sn.
The low-lying position of the 0g$_{7/2}$ orbital is very important
for the M1 and GT properties in this mass region.
Standard Skyrme Hartree-Fock and
Woods-Saxon potential models, whose parameters are determined from the
properties of nuclei near the valley of stability, predict the
0g$_{7/2}$ orbital to be more bound than the 1d$_{5/2}$ orbital by
0.5 to 2.0 MeV.\,$^{13}$

Levels schemes and decay properties of many nuclei have been calculated
and compared to experiment. High-spin yrast levels in $^{104}$Sn,
$^{105}$Sn, $^{106}$Sn, $^{102}$In, $^{103}$In, $^{98}$Cd, $^{100}$Cd,
$^{101}$Cd,
$^{102}$Cd, $^{97}$Ag, $^{98}$Ag, $^{99}$Ag, $^{96}$Rh and $^{97}$Rh
calculated in
the SNB model space
were found to agree with experiment to within a few hundred keV.
(Our results for $^{104}$Sn and $^{106}$Sn are in somewhat better
agreement with experiment than those obtained with the G matrix
approach of Engeland et al..\,$^{14}$)
In addition, the splittings of the low-lying 1/2$^{-}$ and 9/2$^{ + }$
states in the odd-proton nuclei and the 5/2$^{ + }$ and 7/2$^{ + }$
states
in the odd-neutron nuclei are reproduced, and the
closely spaced states in the low-lying odd-odd
multiplets\,$^{5,15}$
are reproduced about as well as the results of previous
calculations.\,$^{6,16}$

We concentrate in this letter on the Gamow-Teller (GT) $\beta^{ + }$
decay
properties of nuclei near $^{100}$Sn. We will compare with recent
experiments and comment on the significance of the predictions
for future experiments. First we discuss the decay of the
even-even N=50 isotones which have been the subject of
several previous theoretical
calculations.\,$^{4,5,6,17}$
In Fig.\ 1 experimental B(GT) values deduced from the
from the $\beta^{ + }$ decay of $^{94}$Ru,\,$^{18}$ $^{96}$Pd\,$^{15}$
and $^{98}$Cd\,$^{5}$ are compared to the SNB
calculation. For purposes of comparison, the theoretical
B(GT) values have been divided by four which represents the
typical overall hindrance of experiment compared to theory.
We will concentrate first on the shape of the GT strength
distribution and then discuss to the origin
of the hindrance. The dashed line represents the
experimental sensitivity limit $-$ that is, a B(GT) of this
value would result in a gamma transition which is too weak
to be observed in the present experiments. The
calculated GT strength distributions are in reasonable agreement with
experiment. The small Q$_{ec}$ window
for the $^{94}$Ru decay allows for only a small fraction of the
GT strength to be observed experimentally. But, by the
time one reaches $^{98}$Cd, the Q$_{ec}$ window is large enough
to allow for most of the calculated strength to be observed
experimentally.
The Skouras and Manakos calculations\,$^{4}$
 obtain
the mean energy of the GT distribution of $^{98}$Cd 0.5 to 1.0
MeV too high compared to experiment. In the present
calculation, the mean energy of
the GT distribution was lowered and brought into
better agreement with experiment
when the proton-neutron TBME were renormalized by the
factor of 0.7 discussed above.

The total GT strength extracted from the $^{98}$Cd decay
experiment is 3.5$^{ + 0.8}_{-0.7}$ compared to a total of
13.4 calculated to lie within the sensitivity limit.
Experiment is thus hindered by about a factor of
$  h_{exp}  $=3.8$^{ + 0.7}_{-0.6}$ compared to theory.
Understanding of
this hindrance is important in general and in particular
for the calculations of the nuclear
double-beta decays\,$^{19}$ which are used to set limits on the
neutrino mass.

In the 0d1s shell nuclei (A=16-40)
one observes a factor of $  h_{high}  $=1/0.6=1.67 hindrance when
experimental GT strengths are compared to those
calculated within the {\it full} 0d1s model space.\,$^{20}$ From
comparison of M1 and GT matrix elements one
can deduce that about two-thirds (in the amplitude) of this comes from
higher-order configuration mixing while one-third comes
from the delta-particle nucleon-hole admixture.\,$^{21}$
Observation of about the same hindrance factor
for the total $\beta^{-}$ strength in heavy nuclei
deduced from (p,n) reactions\,$^{22}$
indicate that the mass dependence of
higher-order and delta admixture effects
is not large, and one may
expect about the same factor of $  h_{high}  $=1.67 to contribute in
the $^{100}$Sn region. This leaves another factor
of $  h_{exp}/h_{high}  $=2.3$\pm$0.4 to be understood.

The calculation
for $^{100}$Sn in the SNB model space is extremely simple $-$
just a single 0g$_{9/2}$ proton hole $-$ 0g$_{7/2}$ neutron
particle final state with a B(GT)=17.8. Instead of this
simple calculation, we show in Fig.\ 1 a calculation within
a 2p2h model space.
The 2p2h model space\,$^{23}$
allows for two-particle two-hole (2p2h) admixture
in the $^{100}$Sn initial state and 2p2h and 3p3h admixture
in the $^{100}$In 1$^{ + }$ final states and thus explicitly
includes the core-polarization
correction calculated in perturbation theory by Towner\,$^{24}$
and Johnstone,\,$^{6}$ as well as some
higher-order terms.
The dimension of the final state is
about 6000, and it is not possible to include more
particle-hole states in the calculation or to carry out a
similar calculation for $^{98}$Cd. The results for $^{100}$Sn
are very interesting. The lowest 1$^{ + }$ state remains
predominantly 1p1h in structure but the strength is
reduced to 80\% of that calculated in the SNB model space.\,$^{25}$
The final states which have a predominantly 2p2h and
3p3h structure
do not start in the spectrum until about 6 MeV in
excitation and carry only a few percent of the total
GT strength. For the analogous calculations in the
0d1s and 0f1p shells,\,$^{26}$ the
simple state and complex states are nearly degenerate in
energy resulting in a spreading of the GT strength
over many states (a large spreading width).
The very different result for $^{100}$Sn
is due to the relative reduction of the residual
interaction compared to the 0g$_{9/2}$-0g$_{7/2}$
spin-orbit splitting and to the fact that both the
0g$_{9/2}$ and 0g$_{7/2}$ orbitals lie next to the Fermi surface.
As has been pointed out,\,$^{27}$
it is Coulomb interaction
which pushes the proton 0g$_{9/2}$ SPE above the neutron
0g$_{7/2}$ SPE and opens up the Q-value window for this
strong GT decay. The 0g hindrance factor
we obtain for $^{100}$Sn of $  h_{0g}  $=1.25 is
smaller than the results obtained in perturbation theory by
Johnstone\,$^{6}$ ($  h_{0g}  $=1.60) but consistent with the
interaction-dependent range given by Towner\,$^{17}$
($  h_{0g}  $=1.29-1.71).

Some Z dependence is expected for the 0g hindrance factor.
The results of Towner and Johnstone for the
ratio $  h_{0g}  $($^{98}$Cd)/$  h_{0g}  $($^{100}$Sn) range from
1.23 to 1.30 and are much less interaction dependent than the
actual range of values given above.
Assuming a ratio of 1.30,
our hindrance factor of $  h_{0g}  $=1.25 for $^{100}$Sn would translate
into a factor of $  h_{0g}  $=1.62 for $^{98}$Cd compared to
$  h_{exp}/h_{high}  $=2.3$\pm$0.4.
We speculate in analogy with the 0d1s and 0f1p shell
calculations,\,$^{26,28}$
that higher-order mixing
between the 0g$_{9/2}$ and 0g$_{7/2}$ orbitals is responsible for the
remaining
hindrance in the $^{100}$Sn region
$-$ such calculations for $^{100}$Sn region
may soon be possible within the
Monte-Carlo shell-model approach.\,$^{28}$ The experimental
hindrance obtained for $^{100}$Sn compared to that of $^{98}$Cd
will be important in deciding which hindrance mechanism is most
important.
[Using $  h  $= 2.09 (=1.67x1.25) and Q$_{ec}$=7 MeV
we calculate T$_{1/2}$($^{100}$Sn)=0.53s.]

Similar calculations have been also performed for the GT decays of
odd-A  and odd-odd nuclei in the vicinity of $^{100}$Sn. Before this
work
the GT-strength distributions for the decays of non even-even
nuclei in the region of $^{100}$Sn were presented only for
$^{93}$Tc and $^{95}$Rh.\,$^{6}$ In this letter
we present as examples the GT strength
distributions obtained for the decays of $^{101}$Sn and
$^{100}$In $-$ the closest
neighbors of $^{100}$Sn.
Over one hundred levels in $^{101}$In are expected to be fed in the
decay
of $^{101}$Sn, see Fig 2a. Most of the strength is found at high
excitation
energies well above the proton separation energy
(S$_{p}$ $\approx$ 1.3 MeV) in the $^{101}$In isotope. This leads to a
beta-delayed proton branching ratio above 40 percent,
and explains
why it was possible at all to detect a few tens of the protons
assigned to the decay of $^{101}$Sn which was produced
in the heavy-ion fusion-evaporation reaction at the cross-section
level of about 2 microbarns and identified at the on-line mass
separator.\,$^{29}$
Using $  h  $=4
we calculate T$_{1/2}$($^{101}$Sn)=1.4s which is
close to the experimental limit
of T$_{1/2}$ $\geq$ 1s.\,$^{2}$

The GT decay of $^{100}$In, which has a theoretical
ground state spin of 6$^{ + }$, is shown in Fig.\ 2b.
Most of the strength is located in a broad symmetric peak
centered at about 6 MeV. In addition, a small side peak at
about 2.5 MeV can be seen. It is interesting to notice the
similarity
of calculated GT distribution for $^{100}$In with the experimental one
obtained for decay of $^{104}$In using Total Absorption
Gamma Spectrometer TAGS.\,$^{30,31}$
(The latter decay cannot be calculated due to the
large number of neutron valence particles.) The TAGS method allows
one, in principle,
to obtain the ``true" GT-distributions even for such complex decays
with high gamma multiplicity and statistical gamma cascades following
beta decay. The $^{104}$In decay is limited by a Q$_{ec}$ value which is
about 2 MeV lower than the
one for $^{100}$In, which results in a cutoff of the
GT-strength at higher excitation energies.
However the theoretical picture for $^{100}$In
resembles the main GT strength features measured already for $^{104}$In.
Using $  h  $=4
we calculate T$_{1/2}$($^{100}$In)=6.8s which is
close to the experimental result
of T$_{1/2}$=5$\pm$1s.\,$^{3}$

In summary, we predict a very simple $\beta^{ + }$ decay mode for
$^{100}$Sn. The
experimental observation of beta-delayed gammas and/or protons will
provide a test of the model, and the hindrance factor obtained for this
decay compared to that of $^{98}$Cd will provide a test of the hindrance
mechanism. The calculated GT decays of $^{101}$Sn and $^{100}$In show
the
importance of being able to measure the total decay energy in a TAGS
experiment.

\vspace{ 24pt}
\noindent
  {\bf Acknowledgments}

\vspace{ 12pt}
Part of this work was carried out during our visit to GSI and we would
like to thank Ernst Roeckl and Wolfgang Noerenberg
for their hospitality during this stay. BAB
would like to acknowledge support from US National Science Foundation
grant numbers PHY-90-17077 and PHY-94-03666 and from the Alexander von
Humboldt foundation.

\clearpage
\noindent
  Caption to Fig.\ 1

\vspace{ 24pt}
Gamow-Teller strength distributions for the even-even N=50
isotones. The theoretical calculations on the left
are compared to experiment on the right. For this comparison
the theory has been divided by a factor of four.\ The amount of
GT strength which lies outside the sensitivity limit and Q$_{ec}$
window is indicated.

\vspace{ 36pt}
\noindent
  Caption to Fig.\ 2

\vspace{ 24pt}
The calculated Gamow-Teller strength distributions for
(a) $^{101}$Sn and (b) $^{100}$In.
\clearpage
\noindent
  {\bf References:}

\vspace{ 24pt}
\begin{enumerate}
\item[1    ]R. Schneider et al., submitted to Z. Phys. A.

\item[2    ]H.Keller et al, contribution to the Int. Meeting
     on ``Nuclear Shapes and Nuclear Structure at Low Excitation Energies",
     Antibes, France, June 20-25,1994; R.Grzywacz, diploma thesis,
     Warsaw University 1993.

\item[3    ]G. Reusen et al, ``New beta-delayed proton precursors
     $^{100}$In and $^{102}$In", to be published.

\item[4    ]L. D. Skouras and P. Manakos, J. Phys. G19, 731 (1993).

\item[5    ]A. Plochocki et al., Z. Phys. A342, 43 (1992).

\item[6    ]I. P. Johnstone, Phys. Rev. C44, 1476 (1991).

\item[7    ]W. F. Piel et al., Phys. Rev. C41, 1223 (1990).

\item[8    ]X. Ji and B. H. Wildenthal, Phys. Rev. C37, 1256 (1988).

\item[9    ]H. Kruse and B. H. Wildenthal, Bull. Am. Phys. Soc.
     27, 725 (1982).

\item[10   ]A. Hosaka, K. I. Kubo and H. Toki, Nucl. Phys. A244,
     76 (1985).

\item[11   ]D. H. Gloeckner and F. J. D. Serduke, Nucl. Phys.
     A220, 477 (1974).

\item[12   ]H. P. Blok et al., Nucl. Phys. A273, 142 (1976).

\item[13   ]S. Kamerdzhiev et al., Z. Phys. A346, 253 (1993);
     G. A. Leander et al., Phys. Rev. C30, 416 (1984).

\item[14   ]T. Engeland et al., Phys. Rev. C48, R535 (1993).

\item[15   ]K. Rykaczewski et al., Z. Phys. A322, 263 (1985).

\item[16   ]H. Grawe, R. Schubart and K. H. Maier, private communication.

\item[17   ]I. S. Towner, Nucl. Phys. A444, 402 (1985).

\item[18   ]P. Graf and H. Munzel, Radiochim. Acta 20, 140 (1987).

\item[19   ]W. C. Haxton and G. J. Stephenson,
     Progress in Particle and Nuclear Physics 12, 409 (1984).

\item[20   ]B. A. Brown and B. H. Wildenthal, Atomic Data and Nuclear
     Data Sheets 33, 347 (1985).

\item[21   ]B. A. Brown and B. H. Wildenthal, Nucl. Phys. A474,
     290 (1987).

\item[22   ]C. D. Goodman and S. D. Bloom, ``Spin Excitations in
     Nuclei", ed. F. Petrovich et al., (Plenum, New York), 143 (1984).

\item[23   ]We omitted the 0h$_{11/2}$ neutron orbital. For the neutron
TBME we
     used a modified surface delta interaction with the two constants chosen
     to approximately reproduce the binding energies of the 0$^{+}$ and 6$^{+}$
     states in $^{104}$Sn obtained in model space SNB.

\item[24   ]A. Arima et al., Adv. Nucl. Phys. 18, 1 (1987);
     I. S. Towner, Phys. Rep. 155, 264 (1987).

\item[25   ]Since the proton separation energy
     for $^{100}$In is estimated to be 1.2 MeV, [G. Audi and A. H. Wapstra
     Nucl. Phys. A565, 1 (1993)], the predicted excitation
     energy of 1.8 MeV for the low-lying 1$^{+}$ state is still low
     enough that it should predominantly gamma decay. Proton decay
     (which would be easier to detect experimentally) would start
     to become important if the excitation energy were about
     3 MeV or higher.

\item[26   ]N. Auerbach et al., Nucl. Phys. A556, 190 (1993).

\item[27   ]I. Hamamoto and H. Sagawa, Phys. Rev. C48, R960 (1993).

\item[28   ]Y. Alhassid et al., Phys. Rev. Lett. 72, 613 (1994).

\item[29   ]E. Roeckl, GSI Nachrichten 9-93 (1993) page 3

\item[30   ]L.Batist et al, GSI Scientific Report 1992,
     GSI-93-1 (1993) p.82

\item[31   ]K.Rykaczewski, in Proc. of 6th Int. Conf. on Nuclei Far From
     Stability and 9th Int. Conf. on Atomic Masses and Fundamental Constants,
     Bernkastel-Kues, Germany 1992, IOP Conf. Ser. 132 (1993) p.517.

\end{enumerate}
\end{document}